# Quantum Oscillations in the Field-induced Ferromagnetic State of MnBi$_{2-x}$Sb$_x$Te$_4$

Qianni Jiang[1], Chong Wang[2], Paul Malinowski[1], Zhaoyu Liu[1], Yue Shi[3], Zhong Lin[1], Zaiyao Fei[1], Tiancheng Song[1], David Graf[4], Xiaodong Xu[1,3], Jiaqiang Yan[5], Di Xiao[2], and Jiun-Haw Chu[1]

[1]Department of Physics, University of Washington, Seattle, WA 98105, USA
[2]Department of Physics, Carnegie Mellon University, Pittsburg, PA 15213, USA
[3]Department of Material Science and Engineering, University of Washington, Seattle, WA 98105, USA
[4]National High Magnetic Field Laboratory, Florida State University, Tallahassee, FL 32306, USA
[5]Materials Science and Technology Division, Oak Ridge National Laboratory, Oak Ridge, TN 37831, USA

**Abstract**

The intrinsic antiferromagnetic topological insulator MnBi$_2$Te$_4$ undergoes a metamagnetic transition in a c-axis magnetic field. It has been predicted that ferromagnetic MnBi$_2$Te$_4$ is an ideal Weyl semimetal with a single pair of Weyl nodes. Here we report measurements of quantum oscillations detected in the field-induced ferromagnetic phase of MnBi$_{2-x}$Sb$_x$Te$_4$, where Sb substitution tunes the majority carriers from electrons to holes. Single frequency Shubnikov-de Haas oscillations were observed in a wide range of Sb concentrations ($0.54 \leq x \leq 1.21$). The evolution of the oscillation frequency and the effective mass shows reasonable agreement with the Weyl semimetal band-structure of ferromagnetic MnBi$_2$Te$_4$ predicted by density functional calculations. Intriguingly, the quantum oscillation frequency shows a strong temperature dependence, indicating that the electronic structure sensitively depends on magnetism.

**Introduction**

The recently discovered magnetic topological material MnBi$_2$Te$_4$ presents a unique platform to study band topology intertwined with magnetic order [1-4]. The crystal structure of MnBi$_2$Te$_4$ (Fig. 1a) consists of van der Waals bonded septuple layers. Each of the septuple layers can be viewed as a natural heterostructure of a magnetic MnTe layer sandwiched by Bi$_2$Te$_3$ topological insulators. The layered structure enables exfoliation to reach the regime of two-dimensional phenomenon. For example, the quantum anomalous Hall effect has been observed in atomically thin flakes [5-8]. The ground state of bulk MnBi$_2$Te$_4$ is a layered antiferromagnet (AFM) ($T_N = 24$ K), in which the individually ferromagnetic Mn-Te layers are coupled antiferromagnetically in the out-of-plane direction (the c-axis), which is also the easy-axis of the moments. Density functional theory (DFT) calculations indicate that the three-dimensional bulk electronic structure in the AFM phase is an antiferromagnetic topological insulator, with an insulating bulk and gapped surface state on the top and bottom surfaces. Although soon after the material realization, angle-resolved photoemission spectroscopy (ARPES) studies of MnBi$_2$Te$_4$ revealed the Dirac surface states predicted by the theory [9-14], controversy remains as to whether the Dirac point is gapped as expected [15].

When a magnetic field is applied along the c-axis, the AFM phase undergoes a meta-magnetic transition, and a ferromagnetic (FM) phase is stabilized at 8 T. Compared with the antiferromagnetic phase the electronic structure in the field-induced FM phase is much less explored. This is partly because the magnetic field and ARPES measurements are incompatible. It has been predicted that the electronic structure in the FM phase of $MnBi_2Te_4$ is an ideal type-II Weyl semimetal, with just two Weyl nodes which are situated on the $k_z$ axis[1,2]. If so, the metamagnetic transition is also a field-induced topological phase transition. Recent scanning tunneling spectroscopy (STS) measurements found that the local density of states is almost unchanged across the AFM-FM transition[16,17], inconsistent with a topological phase transition. However, STS only probes the surface. Clearly, a direct measurement of the bulk electronic structure of FM $MnBi_2Te_4$ is desirable to resolve this issue.

Sb substituted $MnBi_2Te_4$ ($MnBi_{2-x}Sb_xTe_4$) offers the opportunity to study the electronic structure in the ferromagnetic phase via quantum oscillations measurements. The Sb substitution effectively tunes the carriers from electrons to holes while preserving its intrinsic magnetism (19 K $<T_N<$ 24 K)[13,18]. Close to the charge neutrality point, the mobility is enhanced and quantum oscillations have been observed [19]. The measurement of quantum oscillations has been a canonical method to probe the bulk electronic structures of metals and semiconductors. In particular, the doping dependence of the oscillation frequency and effective mass provides strong constraints on the band-dispersion near the band edge.

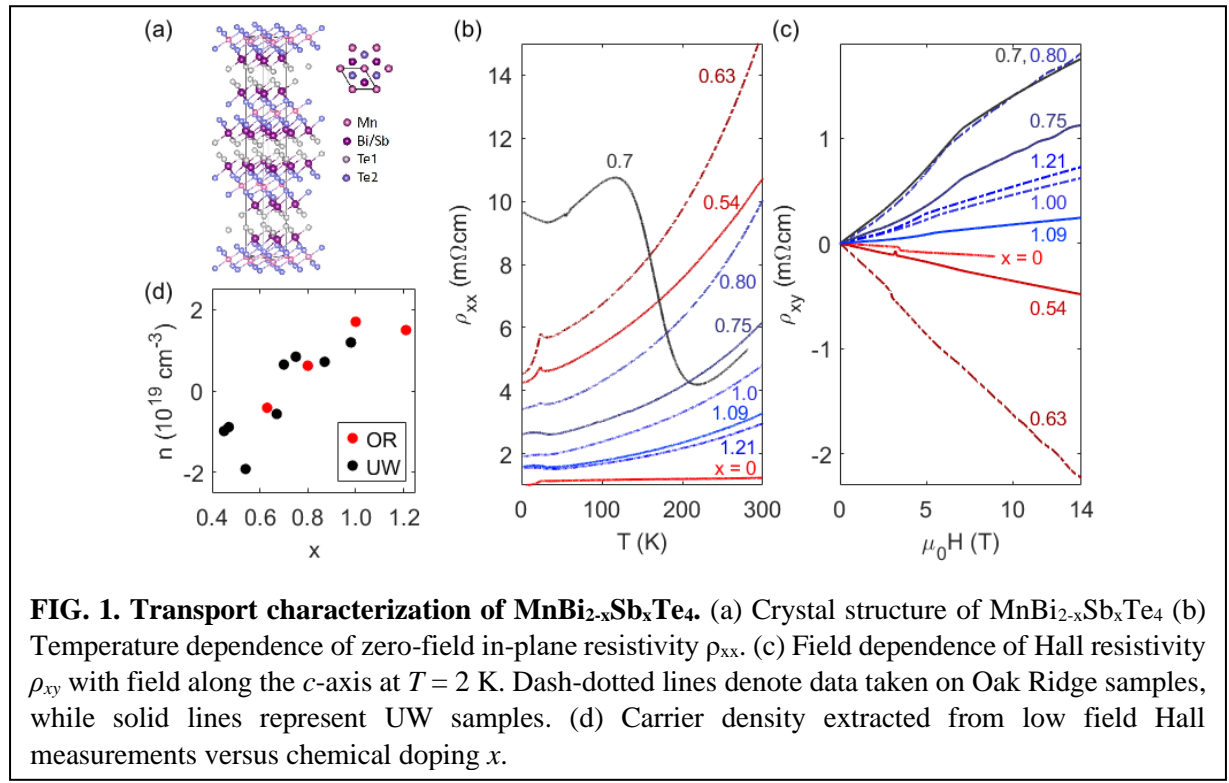

**FIG. 1. Transport characterization of $MnBi_{2-x}Sb_xTe_4$.** (a) Crystal structure of $MnBi_{2-x}Sb_xTe_4$ (b) Temperature dependence of zero-field in-plane resistivity $\rho_{xx}$. (c) Field dependence of Hall resistivity $\rho_{xy}$ with field along the $c$-axis at $T = 2$ K. Dash-dotted lines denote data taken on Oak Ridge samples, while solid lines represent UW samples. (d) Carrier density extracted from low field Hall measurements versus chemical doping $x$.

Here, we report measurements of the Shubnikov de Haas (SdH) oscillations in the field-induced FM phase of $MnBi_{2-x}Sb_xTe_4$ over a wide range of Sb concentrations (0.54 $< x <$ 1.21). By comparing the measured oscillation frequency, carrier density and effective mass with DFT calculations, we find overall reasonable agreement with calculations assuming a rigid band-shift of the Weyl semimetal band-structure in FM MnBi$_2$Te$_4$. Interestingly, the oscillation frequency shows a strong temperature dependence. We can explain this unusual phenomenon as a manifestation of the high sensitivity of the electronic structure to the size of the magnetization in this material.

**Methods**

Single crystals of MnBi$_{2-x}$Sb$_x$Te$_4$ were grown out of a Bi (Sb)-Te flux [18,20]. Initial measurements were performed on samples grown in the Oak Ridge National Lab (OR). After the initial measurements, another batch of the samples were grown at the University of Washington (UW) using the recipe developed by the OR group. The compositions were determined by elemental analysis on a cleaved surface using a Hitachi TM-3000 tabletop electron microscope equipped with a Bruker Quantax 70 energy-dispersive x-ray system for OR samples and a Sirion XL30 scanning electron microscope for UW samples. Both batches of samples exhibited a similar compositional dependence of the physical properties, with a slight offset in the exact composition, likely due to differences in the number of antisite defects[21]. Magnetotransport was carried out in a 36 T series connected hybrid magnet at the National High Magnetic Field Laboratory at Tallahassee, FL (OR samples) and in a 14 T Physical Property Measurement System (PPMS) (UW samples). Magnetotransport measurements were made in a standard four-probe or six-probe contact configuration with current direction in-plane and magnetic field out of the plane (c-axis). To eliminate any effects from contact misalignment, the magneto- and Hall resistivities were symmetrized and anti-symmetrized respectively. Magnetization measurements were made using the vibrating sample magnetometry option of the PPMS.

Density functional theory calculations were performed with the Vienna Ab initio Simulation Package (vasp)[22,23]. A projector-augmented wave [24] method with an energy cut of 269.9 eV

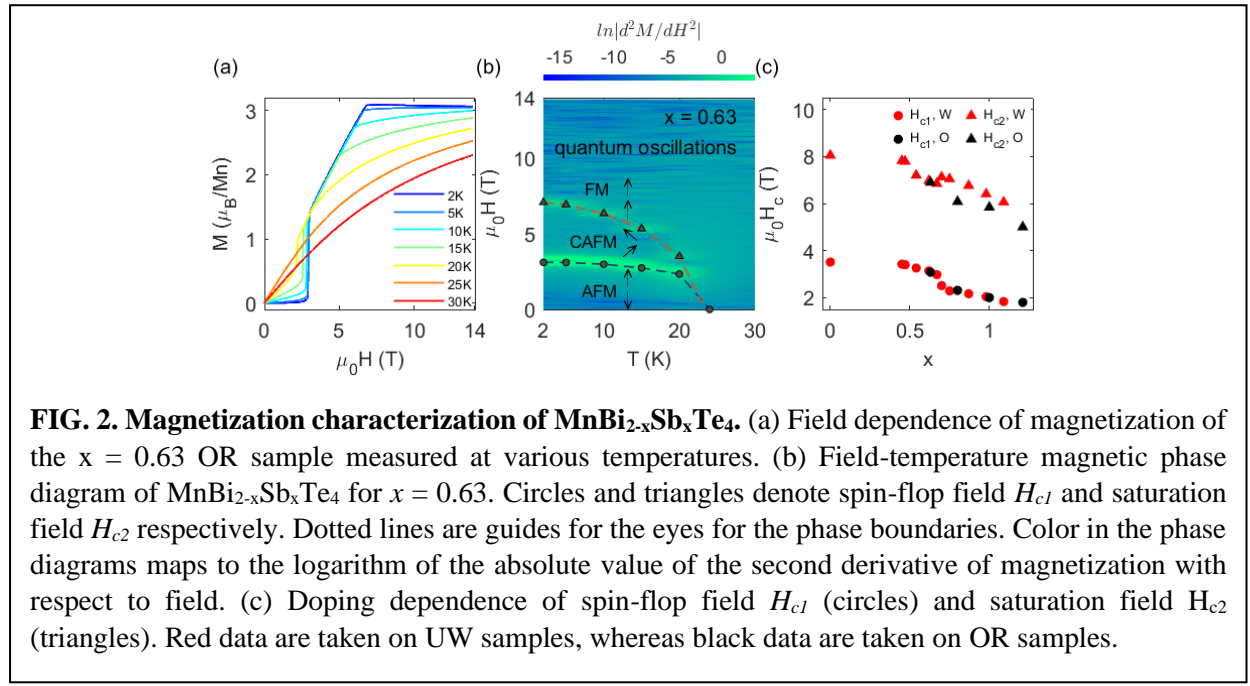

**FIG. 2. Magnetization characterization of MnBi$_{2-x}$Sb$_x$Te$_4$.** (a) Field dependence of magnetization of the x = 0.63 OR sample measured at various temperatures. (b) Field-temperature magnetic phase diagram of MnBi$_{2-x}$Sb$_x$Te$_4$ for $x = 0.63$. Circles and triangles denote spin-flop field $H_{c1}$ and saturation field $H_{c2}$ respectively. Dotted lines are guides for the eyes for the phase boundaries. Color in the phase diagrams maps to the logarithm of the absolute value of the second derivative of magnetization with respect to field. (c) Doping dependence of spin-flop field $H_{c1}$ (circles) and saturation field $H_{c2}$ (triangles). Red data are taken on UW samples, whereas black data are taken on OR samples.

is used to expand Kohn-Sham wave functions. The sampling mesh in reciprocal space is 12×12×4.

The electron-electron interactions are described by modified Becke-Johnson (mBJ) functional[25]. Maximally localized Wannier functions are constructed with the Wannier90 package[26].

**Results**

The magnetic and transport properties of MnBi$_{2-x}$Sb$_x$Te$_4$ have been characterized in previous studies[17,18]. In Fig. 1 and 2 we show representative data for the samples studied in this work. Fig. 1 (b) shows the temperature dependence of the zero-field in-plane electrical resistivity $\rho_{xx}$. For most samples, the resistivity exhibits a metallic behavior, consistent with a degenerately doped semiconductor. For one composition very close to the charge neutrality point ($x = 0.7$), the resistivity shows a non-monotonic temperature dependence. A kink in the resistivity at around 23 to 25 K marks the AFM transition temperature $T_N$. Fig. 1 (c) shows the Hall resistivity $\rho_{xy}$ measured at $T = 2K$. The carrier density is extracted from the weak field Hall coefficient. As shown in Fig. 1 (d), the carrier density changes from $6\times10^{19}$ cm$^{-3}$ electrons to $2\times10^{19}$ holes as the Sb concentration increases, and the system switches from n-type to p-type at a compensation point in the vicinity of $0.6 < x < 0.7$.

Fig. 2 shows the magnetization measurement of a representative sample, $x = 0.63$. As shown in Fig. 2 (a), when the magnetic field is swept along the c-axis at low temperatures, the sample first undergoes an AFM to canted AFM transition, corresponding to the sudden jump of the magnetization at the spin-flop field $H_{c1}$, and then a canted AFM to FM transition at the saturation field $H_{c2}$. Using the magnetization versus field data, we constructed the magnetic phase diagram for x = 0.63, shown in Fig. 2(b). By plotting the second derivative of the magnetization with respect to the field in a log scale, sharp color changes highlight the phase boundaries defined by $H_{c1}$ (circles) and $H_{c2}$ (triangles). As the temperature increases, the FM phase eventually crosses over to the paramagnetic (PM) phase. Within the range of the doping focused on in this study, the Sb substitution only slightly suppresses $T_N$, $H_{c1}$ and $H_{c2}$ without altering the magnetic phase behavior [18]. All the quantum oscillations presented below were observed in the FM-PM crossover regime.

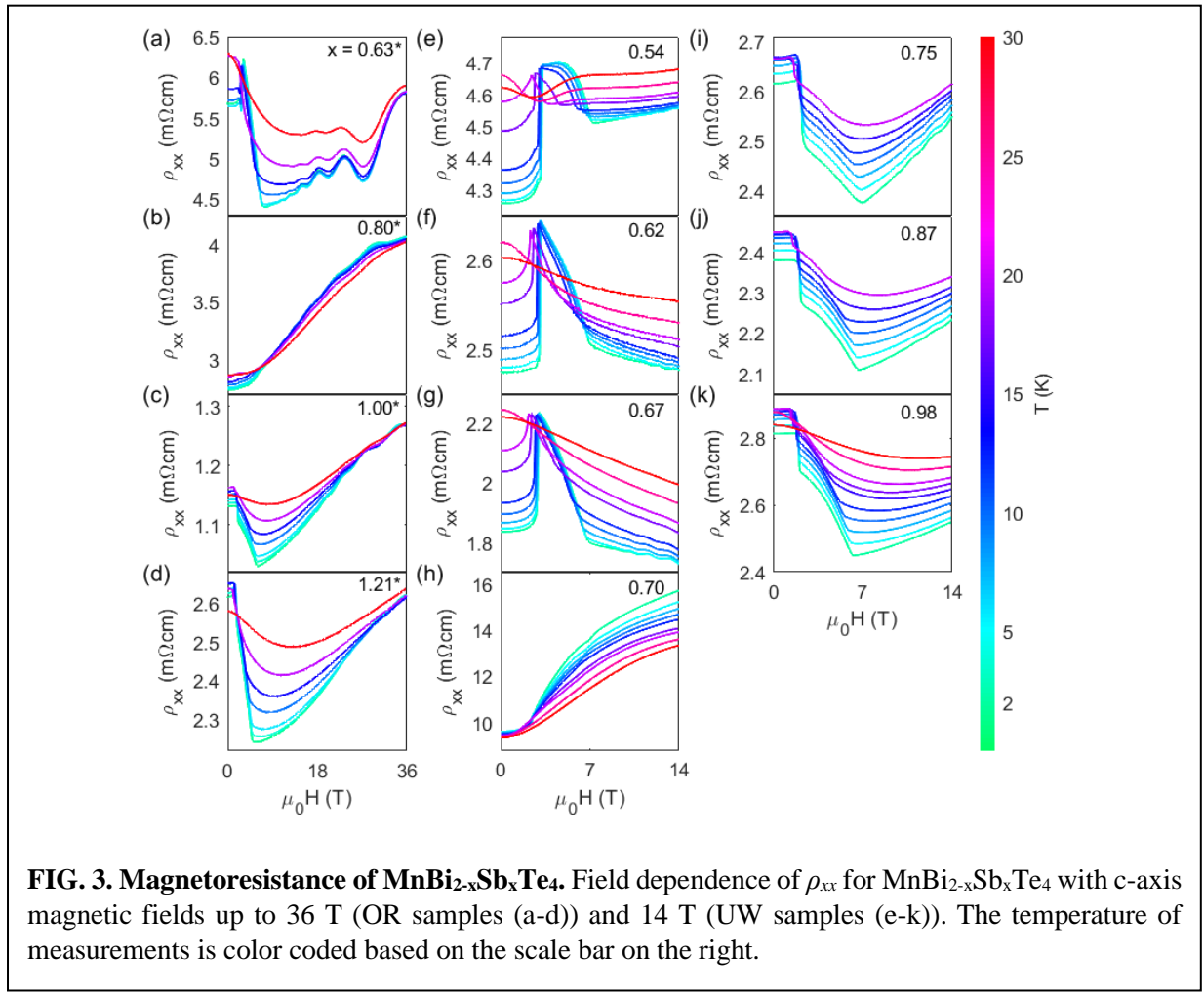

**FIG. 3. Magnetoresistance of MnBi$_{2-x}$Sb$_x$Te$_4$.** Field dependence of $\rho_{xx}$ for MnBi$_{2-x}$Sb$_x$Te$_4$ with c-axis magnetic fields up to 36 T (OR samples (a-d)) and 14 T (UW samples (e-k)). The temperature of measurements is color coded based on the scale bar on the right.

Shubnikov-de Haas oscillations were observed in the transverse magnetoresistance of MnBi$_{2-x}$Sb$_x$Te$_4$. Fig. 3 shows the in-plane longitudninal resistivity $\rho_{xx}$ as a function of c-axis magnetic field. The magnetoresistance (MR) exhibits a strong field dependence. At base temperature, two successive features can be seen in most samples, corresponding to $H_{c1}$ and $H_{c2}$. As the temperature increases, the two anomalies merge and eventually smear out above $T_N$. For all the doping levels shown in the figure, quantum oscillations can be seen once the field surpasses $H_{c2}$. In Fig. 4, the oscillatory part of the MR is plotted against the inverse of the magnetic field. The oscillatory part of MR was extracted by subtracting a fifth-order polynomial background. Using a fast Fourier transform (FFT), we obtained the frequency spectrum for each doping level, plotted in Fig. 5. The oscillation frequency, defined as the peak position in the FFT spectrum at base temperature, is plotted as a function of composition in Fig. 5 (l). Starting from the electron doped side, the oscillation frequency decreases as $x$ increases and reaches 80 T for $x = 0.63$. After crossing the compensation point, the frequency increases monotonically from 36T to 144 T as $x$ increases from 0.7 to 1.21.

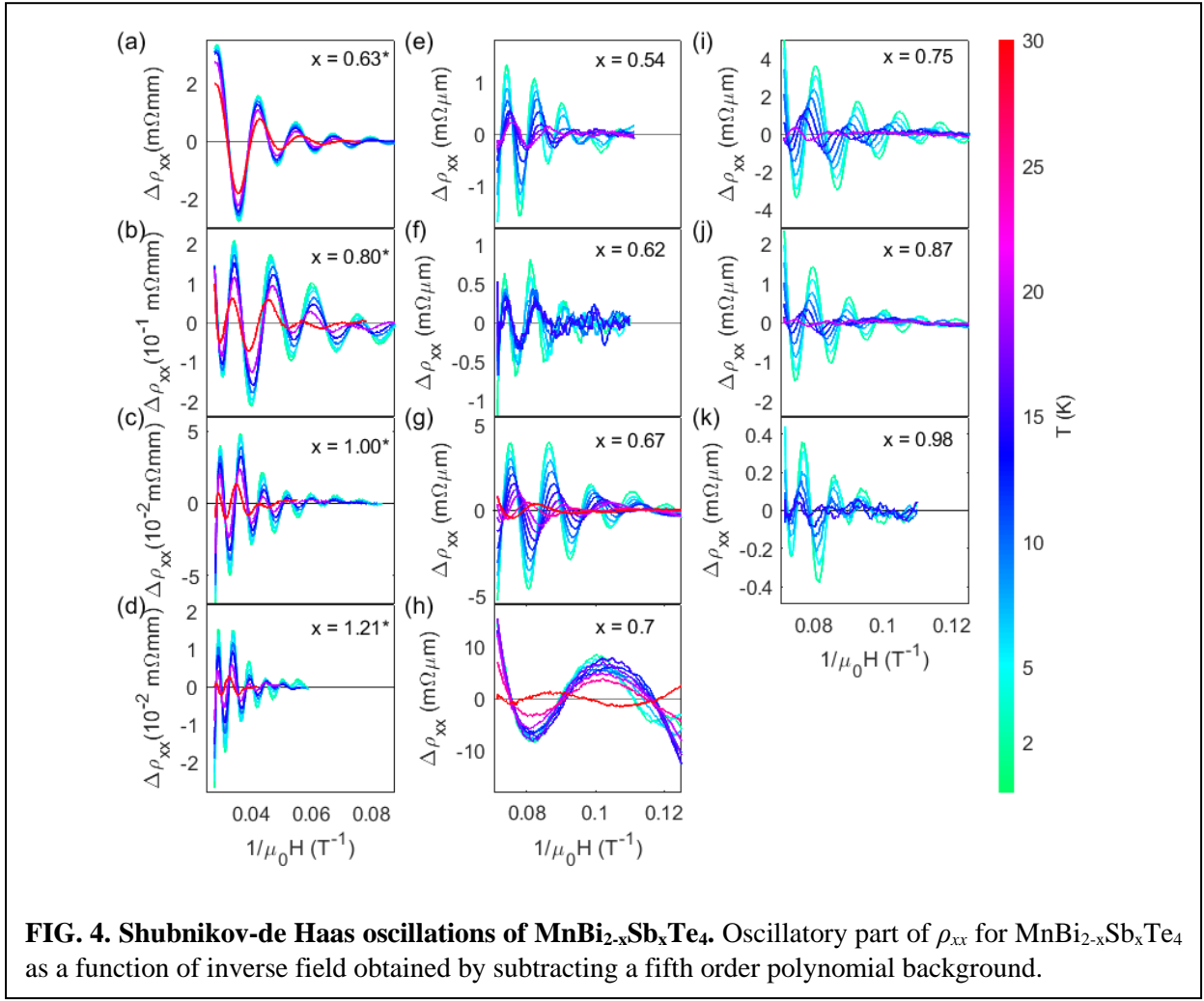

**FIG. 4. Shubnikov-de Haas oscillations of MnBi$_{2-x}$Sb$_x$Te$_4$.** Oscillatory part of $\rho_{xx}$ for MnBi$_{2-x}$Sb$_x$Te$_4$ as a function of inverse field obtained by subtracting a fifth order polynomial background.

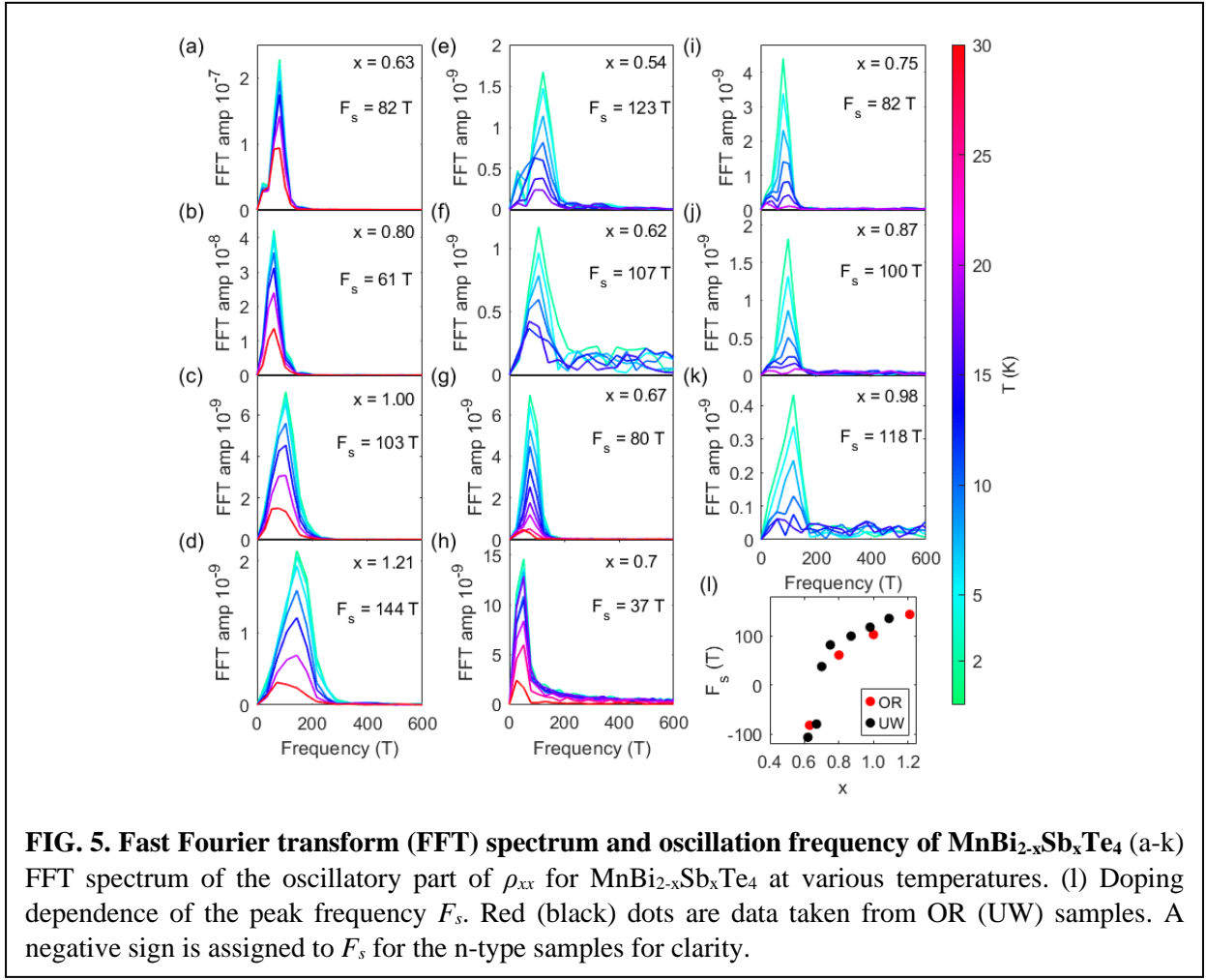

**FIG. 5. Fast Fourier transform (FFT) spectrum and oscillation frequency of MnBi$_{2-x}$Sb$_x$Te$_4$** (a-k) FFT spectrum of the oscillatory part of $\rho_{xx}$ for MnBi$_{2-x}$Sb$_x$Te$_4$ at various temperatures. (l) Doping dependence of the peak frequency $F_s$. Red (black) dots are data taken from OR (UW) samples. A negative sign is assigned to $F_s$ for the n-type samples for clarity.

The quantum oscillation frequency is related to the extremal area of Fermi surface cross section ($A$) projected along the field direction via the Onsager relation $F_s = (\hbar/2\pi e)A$. The compositional dependence of the frequency (Fig. 5(l)) is consistent with the shrinking and expanding of the Fermi surface as the system is tuned across the charge neutrality point. Notice that for a three-dimensional Fermi surface the full angular dependence of the frequency is required to determine the carrier density from its volume. Such a three-dimensional mapping is impossible for MnBi$_{2-x}$Sb$_x$Te$_4$ because the rotation of magnetic field also changes the direction of magnetization, which inevitably changes the electronic structure [27]. In this study, we focus on the electronic structure when the magnetization is aligned to the c-axis by magnetic field.

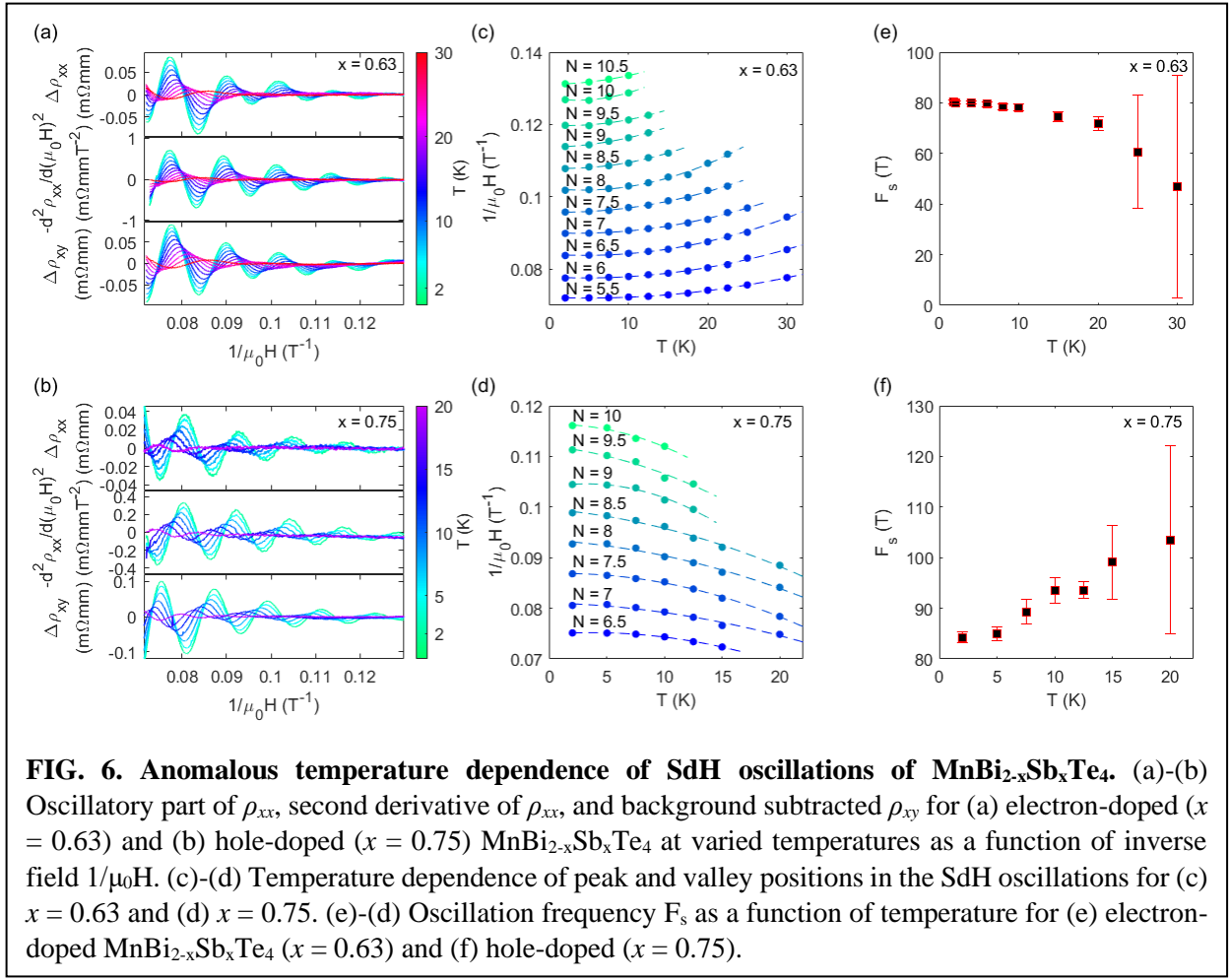

**FIG. 6. Anomalous temperature dependence of SdH oscillations of MnBi$_{2-x}$Sb$_x$Te$_4$.** (a)-(b) Oscillatory part of $\rho_{xx}$, second derivative of $\rho_{xx}$, and background subtracted $\rho_{xy}$ for (a) electron-doped ($x = 0.63$) and (b) hole-doped ($x = 0.75$) MnBi$_{2-x}$Sb$_x$Te$_4$ at varied temperatures as a function of inverse field $1/\mu_0 H$. (c)-(d) Temperature dependence of peak and valley positions in the SdH oscillations for (c) $x = 0.63$ and (d) $x = 0.75$. (e)-(d) Oscillation frequency F$_s$ as a function of temperature for (e) electron-doped MnBi$_{2-x}$Sb$_x$Te$_4$ ($x = 0.63$) and (f) hole-doped ($x = 0.75$).

We next turn to the temperature dependence of the quantum oscillations. The amplitude of quantum oscillations always decays with temperatures due to the smearing of the Fermi-Dirac distribution, but the frequency is usually a constant as a function of temperature. Surprisingly, here, in addition to the thermal damping, we also observed substantial shifts of the peak and valley positions (Fig. 4) as temperature increases. A closer inspection of the FFT spectra (Fig. 5) shows that the oscillation frequency also shifts with temperature. To ensure that this shift is not an artifact due to background subtraction, we extracted the oscillatory data of two representative samples, $x = 0.63$ (electron-doped) and $x = 0.75$ (hole-doped), using three different methods: background subtracted $\rho_{xx}$, second derivative of $\rho_{xx}$ with respect to $H$; and background subtracted Hall resistivity $\rho_{xy}$. The results shown in Fig. 6 (a) and (b) all exhibit the same clear shifts of the peak and valley positions, which are plotted in Fig. 6 (c) and (d). From these we obtained the oscillation frequency by extracting the period in inverse field, which is plotted vs temperature in Fig. 6 (e, f). Interestingly, as the temperature increases the electron-doped sample ($x = 0.63$) shows a decrease in frequency whereas the hole-doped sample ($x = 0.75$) shows an increase in frequency.

In the standard theory, the frequency of quantum oscillations reflects the area of Fermi surface. The Fermi surface may change in principle if the Fermi energy shifts with temperature due to an asymmetric density of states, but such a shift is usually very small. In the cases of Dirac or Weyl semimetals, energy dependent cyclotron mass also leads to a frequency shift as the temperature changes[28]. However, the temperature dependence of oscillation frequency we observed in MnBi$_{2-x}$Bi$_x$Te$_4$ is orders of magnitude larger than the effect caused by band curvature. Hence, such a significant change in frequency suggests instead a modification of the energy bands connected to the temperature-dependent magnetic order. As the temperature increases, the field-induced magnetization decreases due to thermal fluctuations, and the size of the Fermi surface changes accordingly. The opposite trend in electron-doped and hole-doped samples indicates that both the conduction and valence bands move upwards in energy as the temperature increases, so that the electron pocket size decreases while the hole pocket size increases. We note that shifts of the oscillation with temperature have been observed in the magnetic topological semimetal PrAlSi [29], and in the magnetic semimetal CeBiPt [30], suggesting that the phenomenon may be generic to magnetic semimetals.

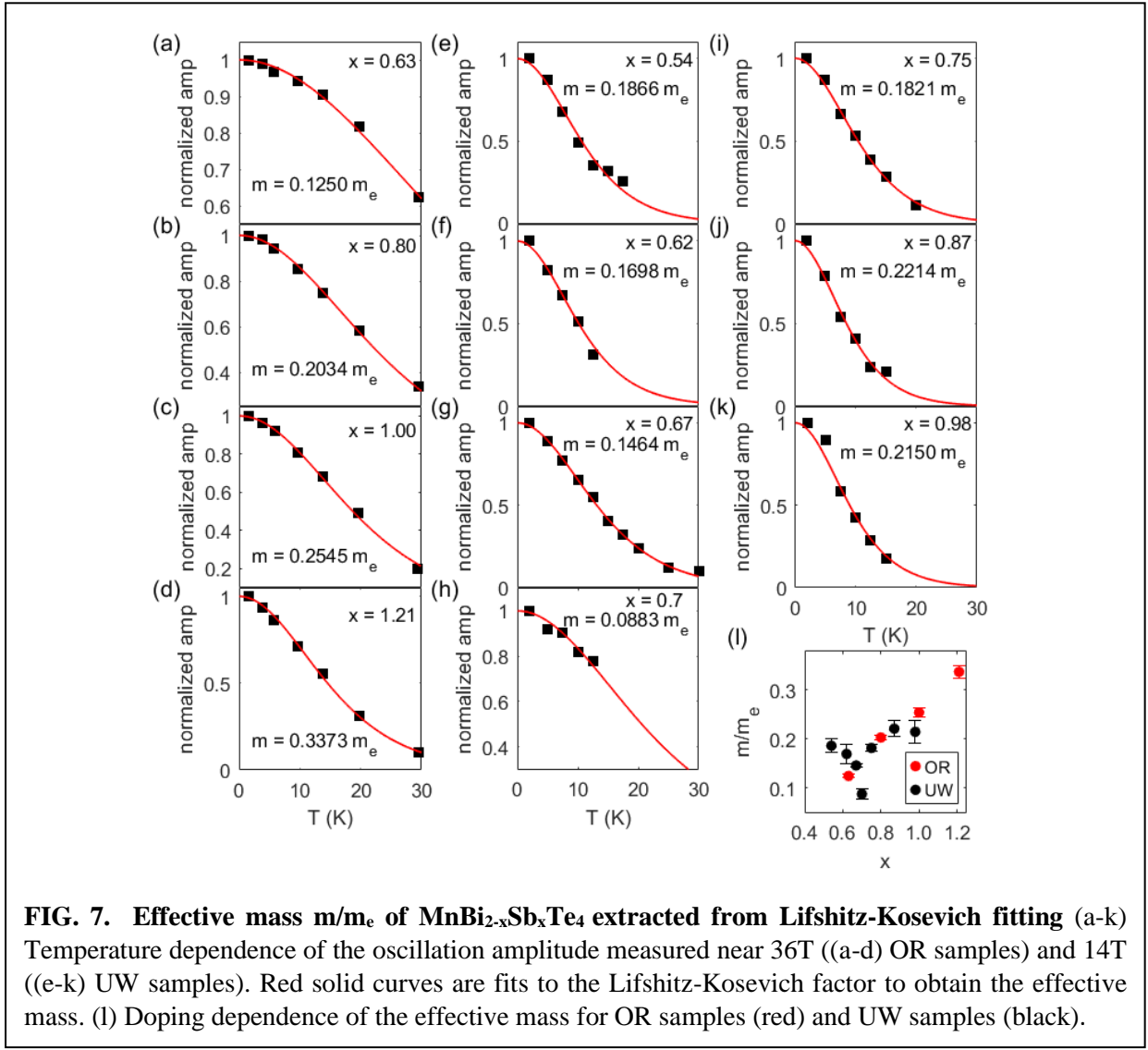

**FIG. 7. Effective mass m/m$_e$ of MnBi$_{2-x}$Sb$_x$Te$_4$ extracted from Lifshitz-Kosevich fitting** (a-k) Temperature dependence of the oscillation amplitude measured near 36T ((a-d) OR samples) and 14T ((e-k) UW samples). Red solid curves are fits to the Lifshitz-Kosevich factor to obtain the effective mass. (l) Doping dependence of the effective mass for OR samples (red) and UW samples (black).

As usual, we proceed to determine the effective mass $m = m^*m_e$ by fitting the temperature dependence of the amplitude of oscillations to the Lifshitz-Kosevich factor,

$$R_T = \frac{\alpha T m^*}{B \sinh(\alpha T m^*/B)}$$

, where $\alpha = 2\pi^2 k_B m_e / e\hbar$. We determine the amplitude by taking the difference between the peak and valley resistances measured at the highest field (~35T for OR samples, ~ 14T for UW samples), where the magnetization is in the fully saturation state and the electronic structure state should be stable. As shown in the Fig. 7, the temperature dependence is well fitted by $R_T$ in all cases. The extracted effective mass shows a strong doping dependence (Fig. 7 (l)), indicating that the energy bands are non-parabolic. In particular, the increase of $m^*$ as the system is doping away from zero-carrier condition is consistent with the massive Dirac dispersion.

To gain further insights, we used DFT to calculate the band structure of ferromagnetic $MnBi_2Te_4$ with the magnetization fully saturated along the c-axis (Fig. 8 (a)). We obtain a Weyl crossing along the Γ-Z direction, consistent with previous studies. As the Fermi energy moves from the heavily electron doped to the heavily hole doped condition, the Fermi surface evolves from a single electron pocket that encloses both Weyl points, to two electron pockets each enclosing a single Weyl point, to a coexistence of one hole pocket and two electron pockets, and finally to one hole pocket that encloses both Weyl points. This is illustrated in Fig. 8 (c) using projections of the Fermi surface onto the $k_y = 0$ plane. The frequencies corresponding to the extremal orbits are plotted in Fig. 8 (b) as a function of Fermi energy. If we map the observed frequencies to the calculation (Fig. 8 (b)), the estimated range of Fermi levels for the samples in this study (shown as the shaded gray region in Fig. 8 (a)) are all outside the range of Lifshitz transitions in which the Fermi surface topology changes, i.e. the Fermi surface is either a single electron or a single hole pocket that encloses both Weyl points.

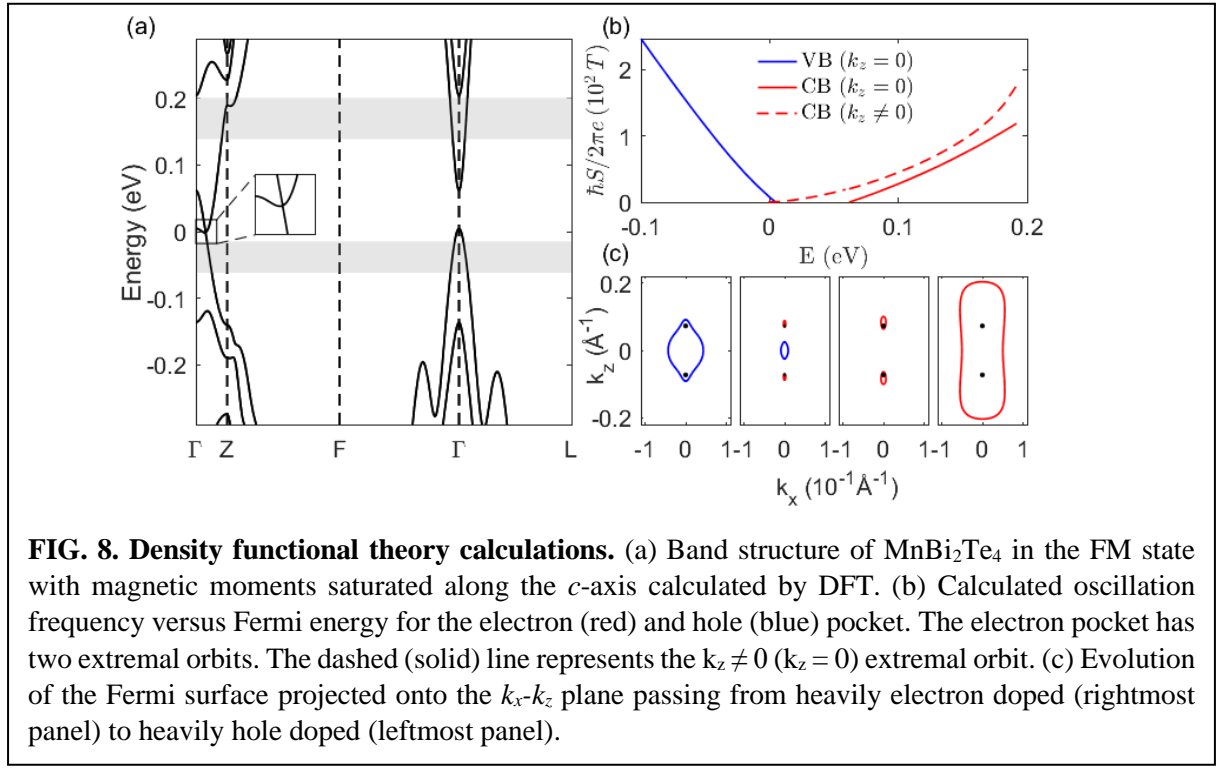

**FIG. 8. Density functional theory calculations.** (a) Band structure of $MnBi_2Te_4$ in the FM state with magnetic moments saturated along the *c*-axis calculated by DFT. (b) Calculated oscillation frequency versus Fermi energy for the electron (red) and hole (blue) pocket. The electron pocket has two extremal orbits. The dashed (solid) line represents the $k_z \neq 0$ ($k_z = 0$) extremal orbit. (c) Evolution of the Fermi surface projected onto the $k_x$-$k_z$ plane passing from heavily electron doped (rightmost panel) to heavily hole doped (leftmost panel).

**Discussion**

The goal of this study was to address two questions about MnBi$_{2-x}$Sb$_x$Te$_4$: (i) Does the electronic structure change across the field-induced meta-magnetic transition? (ii) Is the field-induced FM state an ideal magnetic Weyl semimetal? The strong temperature dependence of the oscillation frequency answers the first question, clearly demonstrating the modulation of the Fermi surface by the magnetism. We can address the second question by making a comparison between the experimentally measured and theoretically calculated effective mass.

The effective mass, $m$, is the derivative of the cyclotron orbit area $A$ with respect to energy, $E$:

$$m = \frac{\hbar^2}{2\pi}\frac{\partial A}{\partial E}$$

The effective mass $m$ as a function of $A$ is determined by the band dispersion $E(\mathbf{k})$, and $A$ is in turn directly related to the oscillation frequency $F_s$ via the Onsager relation, $F_s = (\hbar/2\pi e)A$. Therefore, the experimentally measured variation of $m$ with respect of $F_s$ provides strong constraints on the underlying band dispersion. In Fig. 9 (a), the measured effective mass versus oscillation frequency is plotted alongside the prediction based on the Weyl dispersion (blue and red curves) calculated by DFT and the AFM insulator dispersion (grey dashed curves) based on the $k \cdot p$ model presented in reference [2]. The Weyl dispersion shows a much better agreement with the experimental values compared with the AFM insulator. This is another piece of evidence for a change in electronic structure across the AFM-FM transition. In particular, the calculated values of effective mass of the conduction band match almost perfectly with the experimental data, and the valence band effective mass also falls within the range of measured values, although the experimental doping dependence seems somewhat stronger. We also calculated the carrier density (from the Fermi surface volume) versus oscillation frequency from the Weyl band structures and it too shows good agreement with the experimental data (Fig. 9 (b)).

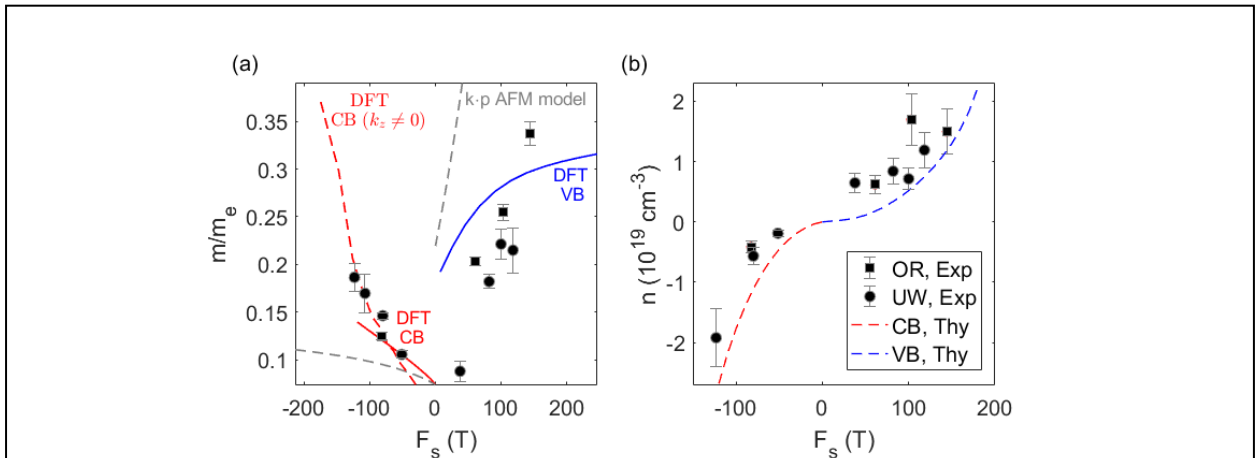

**FIG. 9. Comparison between experiment and theory.** Theoretical predictions (dashed lines) and experimental measurements of the effective mass (a) and carrier density (b) as a function of quantum oscillation frequency. Circles are for UW samples, and squares are for OR samples. In both (a) and (b), red lines represent calculations for the conduction band, and blue lines are calculations for the valence band. In (a), the red dashed line is the calculations for the $k_z \neq 0$ extremal orbit of the conduction band, while the red solid line is for the $k_z = 0$ orbit of the conduction band. The grey dashed lines show calculations from a $k \cdot p$ model of MnBi$_2$Te$_4$ in the AFM state in [2].

Another indicator that has been widely used to determine the topological nature of a band-structure is the phase shift, $\lambda$, of the quantum oscillations[31]. This has three contributions: $\phi_B$, due to geometric phase; $\phi_R$, due to orbital magnetic moment; and $\phi_Z$, due to the Zeeman coupling. It has often been assumed that in 3D metals Dirac-type bands lead to $\phi_B = \pi$ and parabolic bands lead to $\phi_B = 0$. However, a recent comprehensive analysis has shown that $\phi_B$ is in general a continuous quantity and is only fixed to the specific values of 0 and $\pi$ in certain symmetry classes determined by the space group and the type of cyclotron orbits[32]. The cyclotron orbit and the point group symmetry of FM MnBi$_{2-x}$Sb$_x$Te$_4$ do not belong to one of those symmetry classes and therefore we do not expect to observe either $\phi_B = 0$ or $\pi$. Hence the phase shift cannot be used as smoking-gun evidence for Weyl points in this system. For the sake of completeness, however, we present an analysis of the phase shift of the quantum oscillations in the appendix.

We close by noting two reasons for caution. First, Sb substitution may induce effects that differs significantly from a simple rigid energy shift of the band-structure of MnBi$_2$Te$_4$ [13,14]. This question could be addressed by developing new dopants that tune the Fermi level with less change in chemical composition[33]. Secondly, in the DFT calculations the existence of the Weyl nodes in FM MnBi$_2$Te$_4$ is very sensitive to the input parameters, such as lattice constants [1,2]. Future studies on samples with even lower carrier concentrations will help resolve the exact band structures near the putative Weyl points.

**Conclusion**

In summary, we obtained important insight into the bulk electronic structure of the ferromagnetic MnBi$_{2-x}$Sb$_x$Te$_4$ using quantum oscillation measurements. From the temperature dependence of the oscillation frequency, we infer that the electronic structure is sensitive to the magnitude of the magnetization, while in the limit of saturated *c*-axis magnetization we find good overall agreement with band-structure calculations. This lays the foundations for understanding magnetism-induced topological phases in this class of materials.

**Acknowledgements:**

We thank Wonhee Ko, Chau-Xing Liu, Zhongkai Liu, Lexian Yang, Binghai Yan and for helpful discussions. We thank David Cobden for careful reading of the manuscript and valuable suggestions. This work is primarily supported as part of Programmable Quantum Materials, an Energy Frontier Research Center funded by the U.S. Department of Energy (DOE), Office of Science, Basic Energy Sciences (BES), under award DE-SC0019443. Materials synthesis at UW was partially supported by the Gordon and Betty Moore Foundation's EPiQS Initiative, Grant GBMF6759 to JHC. J.Y. acknowledges support from the U.S. Department of Energy, Office of Science, Basic Energy Sciences, Materials Sciences and Engineering Division. A portion of this work was performed at the National High Magnetic Field Laboratory, which is supported by the National Science Foundation Cooperative Agreement No. DMR-1644779, the State of Florida. X.X. and J.H.C. acknowledge the support from the State of Washington funded Clean Energy Institute. J.H.C. also acknowledge the support of the David and Lucile Packard Foundation.

**Appendix: Phase shift of quantum oscillations**

We analyzed the doping dependence of phase shift $\lambda$ by constructing Landau fan diagrams from the quantum oscillations, as shown in Fig. 10. We assigned integer Landau level indices to resistivity peaks, which were shown to coincide with Landau band edges in a study that performed numerical simulation of a minimal magnetic Weyl semimetal [34]. The phase shift $\lambda$ is determined by the following relation:

$$\Gamma = -\frac{1}{2} + \lambda + \delta$$

in which $\Gamma$ is the *y*-intercept of the linear fits to Landau indices *n* vs inverse field, $|\delta|$ is 1/8 with the sign depending on whether the orbit is the minimum of maximum of the Fermi surface. We plot $\Gamma$ as a function of oscillation frequency in Fig. 10 (i). Notice that $\lambda$ contains contribution from Berry curvature, orbital moment and Zeeman coupling, i.e. $\lambda = (\phi_B + \phi_R + \phi_Z)/2\pi$. Their contributions cannot be separated without the analysis of higher harmonics of quantum oscillations, which we did not observe. We also do not expect $\phi_B = 0$ or $\pi$ based on the symmetry of space group and orbits of $MnBi_{2-x}Sb_xTe_4$. Nevertheless, we do observe a rapid change of $\Gamma$ for the lightly hole doped samples, which may be related to the large change of Berry flux as the Fermi level is tuned across Weyl points.

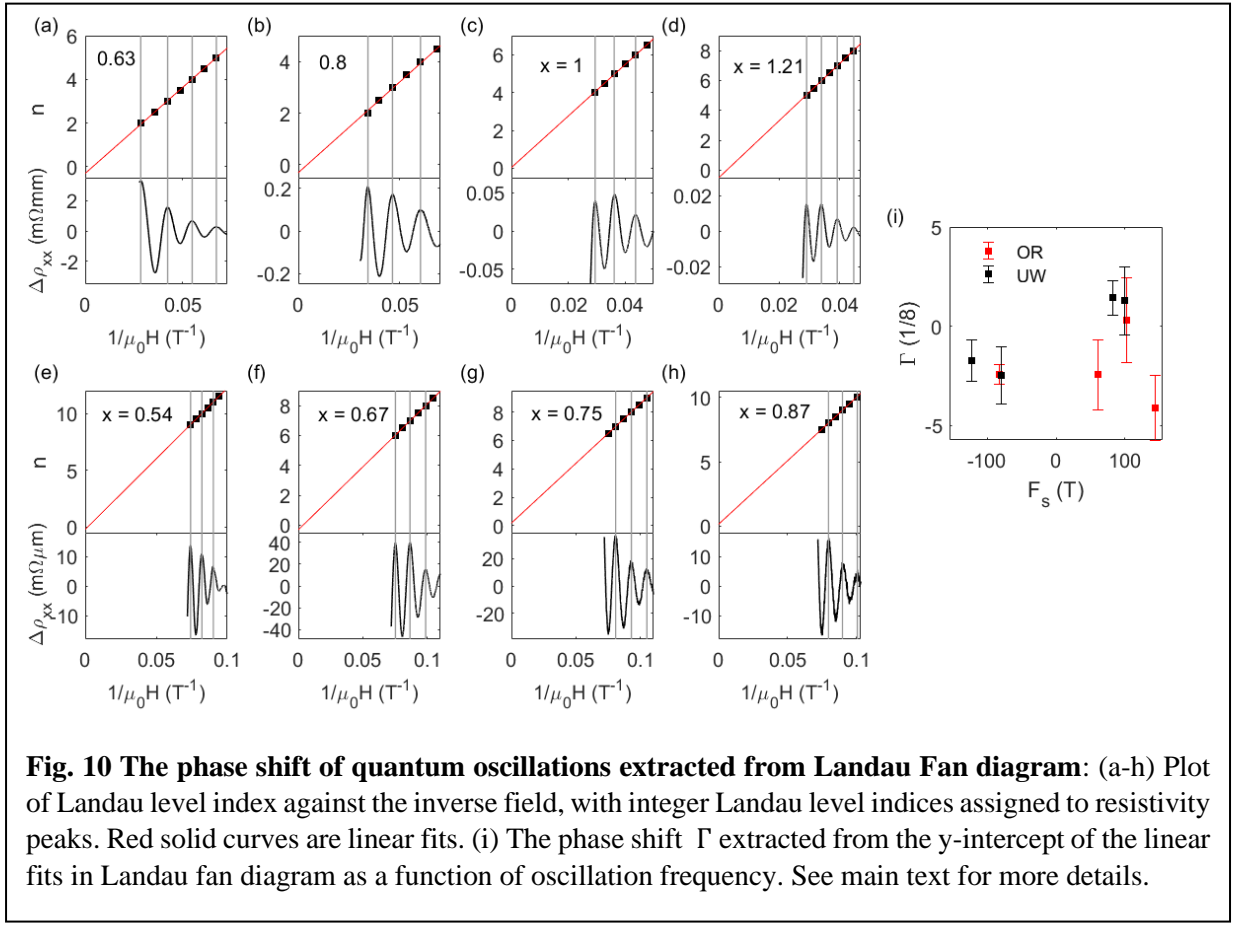

**Fig. 10 The phase shift of quantum oscillations extracted from Landau Fan diagram**: (a-h) Plot of Landau level index against the inverse field, with integer Landau level indices assigned to resistivity peaks. Red solid curves are linear fits. (i) The phase shift $\Gamma$ extracted from the y-intercept of the linear fits in Landau fan diagram as a function of oscillation frequency. See main text for more details.

## References

[1] J. Li, Y. Li, S. Du, Z. Wang, B.-L. Gu, S.-C. Zhang, K. He, W. Duan, and Y. Xu, Science Advances **5**, eaaw5685 (2019).

[2]	D. Zhang, M. Shi, T. Zhu, D. Xing, H. Zhang, and J. Wang, Physical Review Letters **122**, 206401 (2019).
[3]	M. M. Otrokov *et al.*, Nature **576**, 416 (2019).
[4]	Y. Gong *et al.*, Chinese Physics Letters **36**, 076801 (2019).
[5]	C. Liu *et al.*, Nature Materials **19**, 522 (2020).
[6]	Y. Deng, Y. Yu, M. Z. Shi, Z. Guo, Z. Xu, J. Wang, X. H. Chen, and Y. Zhang, Science **367**, 895 (2020).
[7]	J. Ge, Y. Liu, J. Li, H. Li, T. Luo, Y. Wu, Y. Xu, and J. Wang, National Science Review **7**, 1280 (2020).
[8]	D. Ovchinnikov *et al.*, arXiv preprint arXiv:2011.00555 (2020).
[9]	H. Li *et al.*, Physical Review X **9**, 041039 (2019).
[10]	Y. J. Chen *et al.*, Physical Review X **9**, 041040 (2019).
[11]	Y.-J. Hao *et al.*, Physical Review X **9**, 041038 (2019).
[12]	P. Swatek, Y. Wu, L.-L. Wang, K. Lee, B. Schrunk, J. Yan, and A. Kaminski, Physical Review B **101**, 161109 (2020).
[13]	B. Chen *et al.*, Nature Communications **10**, 4469 (2019).
[14]	X.-M. Ma *et al.*, arXiv e-prints, arXiv:2004.09123 (2020).
[15]	D. Nevola, H. X. Li, J. Q. Yan, R. G. Moore, H. N. Lee, H. Miao, and P. D. Johnson, Physical Review Letters **125**, 117205 (2020).
[16]	Y. Yuan *et al.*, Nano Letters **20**, 3271 (2020).
[17]	W. Ko *et al.*, Physical Review B **102**, 115402 (2020).
[18]	J. Q. Yan, S. Okamoto, M. A. McGuire, A. F. May, R. J. McQueeney, and B. C. Sales, Physical Review B **100**, 104409 (2019).
[19]	S. Huat Lee *et al.*, arXiv e-prints, arXiv:2002.10683 (2020).
[20]	J. Q. Yan *et al.*, Physical Review Materials **3**, 064202 (2019).
[21]	Z. Huang, M.-H. Du, J. Yan, and W. Wu, 2020), p. arXiv:2009.07437.
[22]	G. Kresse and J. Furthmüller, Physical Review B **54**, 11169 (1996).
[23]	G. Kresse and D. Joubert, Physical Review B **59**, 1758 (1999).
[24]	P. E. Blöchl, Physical Review B **50**, 17953 (1994).
[25]	A. D. Becke and E. R. Johnson, The Journal of Chemical Physics **124**, 221101 (2006).
[26]	G. Pizzi *et al.*, Journal of Physics: Condensed Matter **32**, 165902 (2020).
[27]	J. Li, C. Wang, Z. Zhang, B.-L. Gu, W. Duan, and Y. Xu, Physical Review B **100**, 121103 (2019).
[28]	C. Guo *et al.*, arXiv preprint arXiv:1910.07608v3 (2020).
[29]	M. Lyu *et al.*, Physical Review B **102**, 085143 (2020).
[30]	J. Wosnitza *et al.*, New Journal of Physics **8**, 174 (2006).
[31]	G. P. Mikitik and Y. V. Sharlai, Physical Review Letters **82**, 2147 (1999).
[32]	A. Alexandradinata, C. Wang, W. Duan, and L. Glazman, Physical Review X **8**, 011027 (2018).
[33]	M.-H. Du, J. Yan, V. R. Cooper, and M. Eisenbach, Advanced Functional Materials **n/a**, 2006516.
[34]	C. M. Wang, H.-Z. Lu, and S.-Q. Shen, Physical Review Letters **117**, 077201 (2016).